\documentclass[showpacs,preprintnumbers,superscriptaddress,prc,nofootinbib,floatfix,twocolumn]{revtex4-1}
\usepackage[unicode]{hyperref}
\usepackage{amsmath}
\usepackage{amsfonts}
\usepackage{amssymb}
\usepackage{graphicx}
\usepackage{dcolumn}
\usepackage{bm}

\usepackage{color} 

\newcommand{\RG}{\textrm{RG}}

\begin{document}

\title{Richardson-Gaudin Configuration-Interaction for nuclear pairing correlations}
\author{Stijn De Baerdemacker}
\email{stijn.debaerdemacker@ugent.be}
\affiliation{Ghent University, Center for Molecular Modeling, Technologiepark 903, 9052 Zwijnaarde, Belgium}
\affiliation{Ghent University, Department of Physics and Astronomy, Krijgslaan 281-S9, 9000 Ghent, Belgium}
\affiliation{Ghent University, Department of Inorganic and Physical Chemistry, Krijgslaan 281-S3, 9000 Ghent, Belgium}
\author{Pieter W.\ Claeys}
\affiliation{Ghent University, Center for Molecular Modeling, Technologiepark 903, 9052 Zwijnaarde, Belgium}
\affiliation{Ghent University, Department of Physics and Astronomy, Krijgslaan 281-S9, 9000 Ghent, Belgium}
\affiliation{Institute for Theoretical Physics, University of Amsterdam, Science Park 904, 1098 XH Amsterdam, The Netherlands}
\author{Jean-S\'ebastien Caux}
\affiliation{Institute for Theoretical Physics, University of Amsterdam, Science Park 904, 1098 XH Amsterdam, The Netherlands}
\author{Dimitri Van Neck}
\affiliation{Ghent University, Center for Molecular Modeling, Technologiepark 903, 9052 Zwijnaarde, Belgium}
\affiliation{Ghent University, Department of Physics and Astronomy, Krijgslaan 281-S9, 9000 Ghent, Belgium}
\author{Paul W. Ayers}
\affiliation{Department of Chemistry and Chemical Biology, McMaster University, Hamilton, Ontario, Canada}
\date{\today}

\begin{abstract}
\begin{description}
\item[Background] The nuclear many-body system is a strongly correlated quantum system, posing serious challenges for perturbative approaches starting from uncorrelated reference states.  The last decade has witnessed considerable progress in the accurate treatment of pairing correlations, one of the major components in medium-sized nuclei, reaching accuracies below the 1\% level of the correlation energy.
\item[Purpose] Development of a quantum many-body method for pairing correlations that is (a) competitive in the 1\% error range, and (b) can be systematically improved with a fast (exponential) convergence rate.
\item[Method] The present paper capitalizes upon ideas from Richardson-Gaudin integrability.  The proposed method is a two-step approach.  The first step consists of the optimization of a Richardson-Gaudin ground state as variational trial state.  At the second step, the complete set of excited states on top of this Richardson-Gaudin ground state is used as an optimal basis for a Configuration Interaction method in an increasingly large effective Hilbert space.
\item[Results] The performance of the variational Richardson-Gaudin (varRG) and Richardson-Gaudin Configuration Interaction (RGCI) method is benchmarked against exact results using an effective $G$-matrix interaction for the Sn region.  The varRG already reaches accuracies around the 1\% level of the correlation energies, and the RGCI step sees an additional  improvement scaling exponentially with the size of the effective Hilbert space.
\item[Conclusions] The Richardson-Gaudin models of integrability provide an optimized complete basis set for pairing correlations.
\end{description}
\end{abstract}

\maketitle

\section{Introduction}
The ground state and low-lying excited states of atomic nuclei are characterized by strong quantum correlations, mainly caused by the strong repulsive core of effective or realistic nucleon-nucleon interactions \cite{bogner:2010}.  This means that many different single-particle configurations are required to give a qualitative account of the low-lying energy physics, giving it a strong multi-reference character.  Fortunately, most of these strong correlations can be nicely understood from symmetry-breaking considerations, often giving rise to an emerging collective behaviour in some symmetry-broken order parameter.  Two famous examples are pairing gaps and quadrupole deformation, associated with $U(1)$ gauge \cite{brink:2005} and $SO(3)$ angular momentum \cite{frauendorf:2001} symmetry breaking respectively.   
Because atomic nuclei are finite-size systems, it is tricky to interpret the emerging broken symmetries as quantum phases from Landau theory, because errors due to quantum fluctuations around the broken symmetry become non-negligible \cite{rowe:1970}.  Therefore, it is important to restore the symmetries, which is typically done by projecting on correct quantum numbers.   Although the projection is not always explicitly performed on the wavefunction, it explains the multi-reference character of the resulting quantum state.  These ideas form the basis of the success of contemporary (beyond) mean-field methods \cite{bender:2003}.

Another approach to capturing strong quantum correlations is by systematically building up the multi-reference character while preserving the symmetry.  This is done in the shell model \cite{heyde:1994} and Coupled Cluster method \cite{hagen:2014}.  Both approaches start from a single reference state, typically the Hartree-Fock (HF) vaccuum, but differ in the way in which other configurations are incorporated.  Whereas the shell model is a typical variational Configuration Interaction (CI) method, Coupled Cluster solves the Schr\"odinger equation in a projective way.  Thanks to the shell structure of atomic nuclei \cite{mayer:1949}, the HF state usually is a good reference state in the vicinity of the (double) shell closures, and the major shell valence space is sufficient to reproduce the degree of collectivity observed experimentally in pairing gaps and quadrupole moments.  Unfortunately,  this is no longer the case when moving towards the mid-shell regions.  Core polarization starts to play a role, and particle-hole excitations across the shell gap give rise to important intruder state configurations \cite{heyde:2011}.  In these cases, it becomes essential to open up the valence space, with the No-Core Shell Model (NCSM) \cite{navratil:2009} as the extreme case where the concepts of core and valence space have been completely eliminated.  This poses a serious computational challenge for CI methods, because the associated Hilbert space scales exponentially\footnote{Technically, the scaling is \emph{combinatorial}.} with the size of the valence space.  It is clear that Moore's law, nor the increase in high-performance computing resources will ever be sufficient to treat medium- to heavy mass nuclei in the NCSM.  Therefore, there is a call for smarter methods than brute force CI in Fock space.  One road to follow is to optimize the basis in which the CI method is constructed.  For instance, the symplectic NCSM \cite{dytrych:2007} answers to this call by constructing a Hilbert space from appropriate irreducible representations of $Sp(3,\mathbb{R})$ that carry the right degree of quadrupole deformation within the basis states, leading to accelerated convergence.  

Nuclear structure physics has a long tradition of building shell-model bases that carry the right degree of (quadrupole) deformation \cite{rosensteel:1977,elliott:1958,nilsson:1995}.  The situation is different for pairing correlations, partially because of the historical success of the symmetry-broken Bardeen-Cooper-Schrieffer (BCS) mean-field state.  However, as nuclear interactions are becoming better constrained and more accurate \cite{bogner:2010}, the mean-field description will no longer be sufficient, and many-body methods are urged to follow along.  Ripoche and collaborators \cite{ripoche:2017} recently proposed a method that combines symmetry projection and CI for pairing correlations, producing ground-state correlation energies with an accuracy of 0.1\% and better.  The core idea is to construct an optimized set of basis states built from the projected BCS state and selected quasi-particle excitations, which are subsequently used in a non-orthogonal CI method.  
 This can be quite well understood physically, because the pairing correlations have already been optimized in the basis states, either at the BCS mean-field level in the strong interaction regime, or at the perturbative particle-hole level in the weak interaction regime.  As such, the approach can be regarded as a natural generalization of the Polynomial Similarity Transformation method (PoST) \cite{degroote:2016}, a many-body method that interpolates between projected BCS theory and pairs Coupled-Cluster Doubles (pCCD) \cite{henderson:2014b}.  The differences between \cite{ripoche:2017} and \cite{degroote:2016} is that the former employs a non-orthogonal CI method, whereas the latter is based on a Coupled Cluster formulation of projected BCS \cite{dukelsky:2016}. 

In the present paper, we adhere to the philosophy of \cite{ripoche:2017}, and put forward an optimized CI basis for pairing correlations in atomic nuclei.  In our case, the optimized basis set will be provided by an integrable Richardson-Gaudin model \cite{dukelsky:2004a}.  It is worth remarking that the previously mentioned methods \cite{ripoche:2017,degroote:2016} performed their test calculations on the Richardson Hamiltonian \cite{richardson:1963}, consisting of an arbitrary\footnote{Typically, one chooses an equidistant energy spectrum, the so-called picket-fence model.} single-particle spectrum with a level-independent pairing interaction\footnote{Other names in the literature include: reduced BCS, level-independent BCS, $s$-wave or Richardson-Gaudin Hamiltonian.}.  The choice for this form of test Hamiltonian is legitimate.  Not only has the Richardson Hamiltonian been put forward as a schematic model to capture pairing correlations in atomic nuclei \cite{bohr:1958}, it was also shown to be \emph{exactly solvable} \cite{richardson:1963} by means of a Bethe Ansatz, turning it indeed into an ideal benchmark for other methods.  The integrability of the Hamiltonian was proven afterwards \cite{cambiaggio:1997}, classifying it within the family of the Gaudin magnets \cite{gaudin:1976,ortiz:2005}.  Integrable models come in many flavors \cite{bethe:1931,lieb:1963,richardson:1963,lieb:1968,gaudin:1976,haldane:1988}, and they have proven particularly useful for elucidating the structure of non-perturbative strongly correlated quantum systems.  Thanks to the Bethe Ansatz structure, the computational cost of obtaining the exact eigenstates and derived observables comes at a polynomially scaling cost, which needs to be appreciated with respect to the exponential cost of conventional exact CI methods.  In the present paper, we will not use these features for the purpose of modeling realistic pairing Hamiltonians (see \cite{rombouts:2004,dukelsky:2011,debaerdemacker:2014} for examples of this), but rather to provide an optimized framework in which to treat realistic pairing Hamiltonians in a CI sense.  In the following sections, we will elaborate on what we will call the Richardson-Gaudin CI method (RGCI).
\section{Richardson-Gaudin model}\label{section:rg}
We will work in the framework of the spherical shell model in the present paper, which assigns to each of the $L$ single particle levels $k$ a unique set of good quantum numbers ($s_kl_kj_km_k\tau_k$), respectively denoting the spin, angular momentum, total angular momentum with its projection, and isospin projection.  For notational reasons, we will refer to the set of quantum numbers and the level itself as $k$.  The Richardson Hamiltonian \cite{richardson:1963} is given by
\begin{equation}\label{rg:hamiltonian}
H_{\RG}=\sum_{k=1}^L\eta_k \hat{n}_k+g\sum_{i,k=1}^L\hat{S}^\dag_k \hat{S}_i,
\end{equation}
with $\eta_k$ the $\Omega_k=2j_k+1$ fold degenerate single-particle energies, and $g$ the level-independent pairing interaction.  Pairing in the spherical shell model happens at the level of the total angular momentum \cite{heyde:1994}, which leads to the definition of pair creation and annihilation operators
\begin{equation}\label{rg:paircreationannihilation}
\hat{S}_k^\dag=\sum_{m_k>0} \hat{a}^\dag_{m_k}\hat{a}^\dag_{\bar{m}_k},\quad \hat{S}_k=(\hat{S}_k^\dag)^\dag=\sum_{m_k>0} \hat{a}_{\bar{m}_k}\hat{a}_{m_k},
\end{equation}
where we have only indicated the index over which the summation runs.  The bar notation $\bar{m}_k$ denotes the time-reversed partner of $m_k$, with a phase correction $a^\dag_{j_k m_k}=(-)^{j_k-m_k}a^\dag_{j_k -m_k}$ in order to respect good angular momentum tensorial properties.  With this notation, the particle-number operators can be written as
\begin{equation}\label{rg:numberoperator}
\hat{n}_k=\sum_{m_k>0}(\hat{a}_{m_k}^\dag\hat{a}_{m_k}+\hat{a}_{\bar{m}_k}^\dag\hat{a}_{\bar{m}_k}),
\end{equation}
again only summing over the relevant index.  It is convenient to introduce the seniority quantum number $v_k$, which counts the number of particles that are not paired as in Eq.\ (\ref{rg:paircreationannihilation}), and the related quasispin  pairing quantum number $d_k=\frac{1}{4}\Omega_k-\frac{1}{2}v_k$, which denotes (half of) the maximum allowed number of pairs in a level \cite{talmi:1993}.

The Hamiltonian (\ref{rg:hamiltonian}) supports a complete set of Bethe Ansatz wavefunctions of the form
\begin{equation}\label{rg:state}
|\vec{\eta},\vec{x}\rangle = \prod_{\alpha=1}^N\left(\sum_{k=1}^L\frac{\hat{S}_k^\dag}{2\eta_i-x_\alpha}\right)|\theta\rangle,
\end{equation} 
with $\vec{\eta}$ the set of single-particle energies $\eta_k$ ($k=1\dots L$), $\vec{x}$ the set of rapidities $x_{\alpha}$ ($\alpha=1\dots N$) and $N$ the number of pairs in the system.  The state $|\theta\rangle$ is the pair vacuum state, meaning that it contains no paired particles.  A state of the form (\ref{rg:state}) is only an eigenstate of the Hamiltonian (\ref{rg:hamiltonian}) provided the rapidities form a solution to the set of Richardson-Gaudin (RG) equations
\begin{equation}\label{rg:equations}
\frac{1}{2g}+\sum_{k=1}^L\frac{d_k}{2\eta_k-x_{\alpha}}-\sum_{\beta\neq\alpha}^N\frac{1}{x_\beta-x_\alpha}=0,
\end{equation}
for all $\alpha=1\dots N$.  This is a strong result because the diagonalisation of the Hamiltonian (\ref{rg:hamiltonian}) in the conventional basis scales combinatorially \cite{volya:2001}, whereas the RG equations scale linearly with the number of pairs involved.  As soon as the RG equations have been solved, the energy of the associated eigenstate (\ref{rg:state}) is readily given by
\begin{equation}
E=\sum_{\alpha=1}^N x_\alpha + \sum_{k=1}^L\eta_k v_k,
\end{equation}
giving an interpretation of pair energy to the rapidities $x_\alpha$.  Another powerful theorem of integrability is Slavnov's theorem \cite{zhou:2002}, related to the evaluation of wavefunction overlaps $\langle\vec{\eta},\vec{y}|\vec{\eta},\vec{x}\rangle$ with both $|\vec{\eta},\vec{x}\rangle$ and $|\vec{\eta},\vec{y}\rangle$ states of the form (\ref{rg:state}), but not necessarily both eigenstates of (\ref{rg:hamiltonian}).  Slavnov's theorem states that the overlap reduces to the evaluation of a determinant
\begin{equation}
\langle\vec{\eta},\vec{y}|\vec{\eta},\vec{x}\rangle=\frac{\prod_{\alpha,\beta\neq\alpha}^N(y_\beta-x_\alpha)}{\prod_{\alpha<\beta}^N(y_\beta-y_\alpha)(x_\alpha-x_\beta)}\det S(\vec{\eta},\vec{x},\vec{y}),
\end{equation}  
with the matrix elements in the Slavnov determinant given by
\begin{align}
S(\vec{\eta},\vec{x},\vec{y})_{\alpha\beta}=&\frac{y_\beta-x_\beta}{y_\alpha-x_\beta}\left[\sum_{k=1}^L\frac{2d_k}{(2\eta_k-y_\alpha)(2\eta_k-x_\alpha)}\right.\notag\\
&\quad\left.-\sum_{\gamma\neq\alpha}^N\frac{2}{(y_\alpha-y_\gamma)(x_\beta-y_\gamma)}\right],
\end{align}
provided at least $|\vec{\eta},\vec{y}\rangle$ is an eigenstate of the RG Hamiltonian (\ref{rg:hamiltonian}).  We refer to the eigenstates as being \emph{on-shell}, opposed to the \emph{off-shell} states of the form (\ref{rg:state}) that are not eigenstates of an integrable Hamiltonian (\ref{rg:hamiltonian}).  The power of Slavnov's theorem can again be appreciated by confronting it with the conventional way of calculating these overlaps, which is done by explicitly expanding the state in the exponentially scaling Hilbert space and summing over all possible coefficients.  The construction of determinant expressions for the overlaps of Bethe Ansatz states in the Richardson-Gaudin model has been an active research topic in the past decade \cite{zhou:2002,faribault:2008,gorohovsky:2011,faribault:2012,claeys:2015a,faribault:2016,faribault:2017}, giving rise to many different determinant representations which are all interconnected \cite{claeys:2017a}.  For the purpose of this paper, it suffices to note that such computationally facile expressions exist.  We refer to recent papers \cite{claeys:2017a,claeys:2017b} for more technical details.
\section{Richardson-Gaudin Configuration Interaction}
The purpose of this paper is to find the ground state and low-lying excited states of an \emph{arbitrary} pairing Hamiltonian of the form
\begin{equation}\label{rgci:hamiltonian}
H_{\lnot\RG}=\sum_{k=1}^L\varepsilon_k\hat{n}_k+\sum_{i,k=1}^L V_{ik} \hat{S}_i^\dag \hat{S}_k
\end{equation} 
by means of an optimized CI scheme in an RG basis.  The pair scattering matrix $V$ can be arbitrary, and is therefore not constrained by any integrability condition.  The notation $\lnot$RG is introduced to emphasize that the Hamiltonian (\ref{rgci:hamiltonian}) is \emph{not} (necessarily) RG integrable.

\begin{figure}[!htb]
	\includegraphics{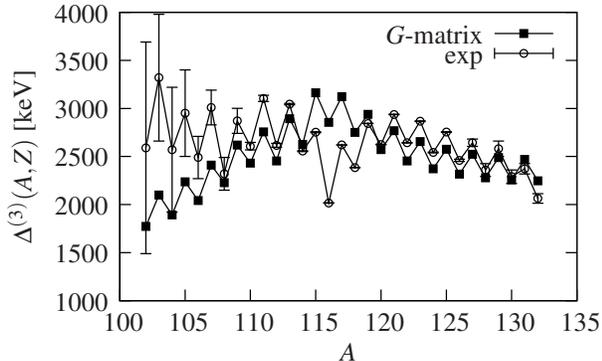}
	\caption{3-point neutron pairing gaps $\Delta^{(3)}$ (\ref{rgci:3pointgaps}) calculated from an effective $G$-matrix interaction (See Table \ref{table:interactionparameters} and \cite{zelevinsky:2003}), and compared with experimental values \cite{audi:2003}.}\label{figure:rgci:gmatrixvsexperiment}
\end{figure}

To fix ideas, we will use an effective pairing interaction obtained from a $G$-matrix construction for the Sn isotopes in the neutron valence shell $A=100-132$ ($g_{\frac{7}{2}},d_{\frac{5}{2}},s_{\frac{1}{2}},h_{\frac{11}{2}},d_{\frac{3}{2}}$) \cite{holt:1998,zelevinsky:2003}.  The specific values of the pairing interaction can be found in \cite{zelevinsky:2003}, and are also listed in Table \ref{table:interactionparameters} for quick reference.  
\begin{table}[!htb]
\begin{tabular}{l|ccccc}
\hline
  & $g_{\frac{7}{2}}$ & $d_{\frac{5}{2}}$ & $s_{\frac{1}{2}}$ & $h_{\frac{11}{2}}$ & $d_{\frac{3}{2}}$ \\
\hline
$\Omega_k$ & $8$ & $6$ & $2$ & $12$ & $4$ \\
\hline
$\varepsilon_k$  & $-6.121$ & $-5.508$ & $-3.891$ & $-3.778$ & $-3.749$ \\
\hline
   $g_{\frac{7}{2}}$  & $-0.2463$ & $-0.1649$ & $-0.1460$ & $-0.2338$ & $-0.1833$ \\
   $d_{\frac{5}{2}}$  &           & $-0.2354$ & $-0.1995$ & $-0.2250$ & $-0.3697$ \\
   $s_{\frac{1}{2}}$  &           &           & $-0.7244$ & $-0.1741$ & $-0.2486$ \\
   $h_{\frac{11}{2}}$ &           &           &           & $-0.1767$ & $-0.1762$ \\
   $d_{\frac{3}{2}}$  &           &           &           &           & $-0.2032$ \\
\hline
\end{tabular}
\caption{Pairing interaction parameters $\varepsilon_k$ and $V_{ik}$ in the Hamiltonian (\ref{rgci:hamiltonian}) for Sn isotopes from a $G$-matrix formalism \cite{holt:1998,zelevinsky:2003}.  All energies are measured in MeV.}\label{table:interactionparameters}
\end{table}
This is an ideal benchmark system for multiple reasons.  First, the pairing strength is known to be very stable in the Sn isotopes, with a slight experimentally observed decrease around the neutron number $64$ subshell closure \cite{jungclaus:2011,morales:2011,debaerdemacker:2014}.  Second, the dimensions of the pairing Hamiltonian (\ref{rgci:hamiltonian}) are rather limited for this shell, so a comparison with exact results from conventional exact CI \cite{volya:2001} remains possible.  To illustrate the performance of the effective interaction (Table \ref{table:interactionparameters}) with respect to experimental values, calculated 3-point neutron pairing gaps, derived from nuclear binding energies $BE(A,Z)$ \cite{bohr:1998},
\begin{align}\label{rgci:3pointgaps}
\Delta^{(3)}(A,Z)=(-)^A[&BE(A,Z)-2BE(A-1,Z)\notag\\
&+BE(A-2,Z)],
\end{align}
are compared to experimental values in Fig.\ \ref{figure:rgci:gmatrixvsexperiment}. 
Experimental values are taken from \cite{audi:2003}.  For the purpose of this paper, it is sufficient to note the qualitative agreement between the $G$-matrix results and experimental values within their errorbars, pointing out that the effective interaction \cite{holt:1998,zelevinsky:2003} is indeed a realistic and non-integrable interaction for pairing correlations.  As such, this interaction will be used solely to test the RGCI method, as is done before with other methods like \cite{sambataro:2012,sambataro:2013}, and we will refrain explicitly from making further comparison with experimental data.
Our approach consists of two consecutive steps.
\begin{enumerate}
\item In a first step, we optimize the basis.  This is done by means of a variational Richardson-Gaudin (varRG) calculation, as has been done before in the context of  quantum chemistry \cite{tecmer:2014,johnson:2015} and integrability-breaking quantum dots \cite{claeys:2017b}.  Because of the integrability of the underlying Richardson-Gaudin model (\ref{rg:hamiltonian}), the variationally obtained state not only gives an approximation of the ground state, but also a complete set of orthogonal basis states, used in the consecutive step.  
\item In the second step, the actual RGCI step, the non-integrable Hamiltonian of interest (\ref{rgci:hamiltonian}) is diagonalized in an increasingly large basis set until convergence is obtained.  This step is very much related to other CI methods acting in a basis of on-shell integrable states, such as the Truncated Space Approximation \cite{yurov:1990,yurov:1991,james:2017} which has been used to diagonalize perturbed integrable quantum field theories in one dimension.  In the present paper, the use of an optimized Richardson-Gaudin basis is key.
\end{enumerate}
Both steps will be discussed in more detail in the following subsections.  Basically, the method is an adaptation of traditional HF+CI methods.  In these methods, an optimal single-particle Hartree-Fock product state is obtained first.  This state then defines a Fock Hilbert space in which residual interactions can be systematically included until convergence.  Again, the main difference in this paper is that the HF state is replaced by a variational Richardson-Gaudin state, already incorporating collective pairing correlations in the initial step.
\subsection{Variational Richardson-Gaudin}
The objective function in the varRG method is the energy functional
\begin{equation}\label{rgci:varrg:functional}
E[\vec{\eta},\vec{x}]=\frac{\langle\vec{\eta},\vec{x}|H_{\lnot\RG}|\vec{\eta},\vec{x}\rangle}{\langle\vec{\eta},\vec{x}|\vec{\eta},\vec{x}\rangle},
\end{equation}
in which the state $|\vec{\eta},\vec{x}\rangle$ (\ref{rg:state}) is used as the trial wavefunction with the additional constraint that it is on-shell, i.e.\ the set of variational parameters $\{\vec{\eta},\vec{x}\}$ satisfy the RG equations (\ref{rg:equations}).  This constraint is required to benefit from the favourable computational scaling provided by Slavnov's theorem and its corrolaries when evaluating the energy expectation value.  Consequently, the variational procedure occurs effectively on a manifold over $\vec{\eta}$ and $g$, as the rapidities are coupled to the single-particle energies  $\vec{\eta}$ via $g$ in the RG equations (\ref{rg:equations}).  We denote the optimal values of $\vec{\eta}$ and $g$ by
\begin{equation}\label{rgci:varrg:optimalparameters}
\{\vec{\eta}_0,g_0\}=\arg\min_{\{\vec{\eta},\vec{x}(\vec{\eta},g)\}}(E[\vec{\eta},\vec{x}]),
\end{equation}
in which we have encoded the implicit dependency of the rapidities $\vec{x}$ on the single-particle states $\vec{\eta}$ and $g$ via the RG equations (\ref{rg:equations}) in the notation $\vec{x}(\vec{\eta},g)$.

The on-shell requirement complicates the variational procedure, because it is important to select the proper manifold of RG eigenstates on which to perform the variational optimization \cite{claeys:2017b}.  Whenever the non-integrable Hamiltonian $H_{\lnot\RG}$ (\ref{rgci:hamiltonian}) is ``close'' to a Richardson Hamiltonian  $H_{\RG}$ (\ref{rg:hamiltonian}), the ground state will be well approximated by the ground state of the corresponding RG Hamiltonian.  In contrast, this is no longer the case when the integrability-breaking terms in $H_{\lnot\RG}$ are large, for which the optimal variational state lives on the manifold of an \emph{excited} Richardson state.  Fortunately, it is well known that attractive pairing Hamiltonians give rise to (collective) Cooper pair formation in the ground state \cite{cooper:1956}, which has a clear-cut correspondence with the ground-state characterisation of the Richardson Hamiltonian \cite{sambataro:2007,debaerdemacker:2012b}.  Therefore, it is safe to assume that these attractive pairing Hamiltonians will be sufficiently close to a RG Hamiltonian (\ref{rg:hamiltonian}).  This was confirmed by our exploratory calculations \cite{claeys:2017b}, in which the ground state of a non-integrable pairing Hamiltonian with random attractive pairing interactions was indeed found to lie on the ground state manifold of the corresponding integrable RG Hamiltonian.  Moreover, the ground state of $H_{\lnot\RG}$ could be rather well approximated by simply replacing the non-integrable pairing interaction by its average.   We will act along the same lines in the present manuscript, and perform the variational calculation in the first step of the procedure only over the parameter $g$, keeping the single-particle parameters fixed as the single-particle energies in $H_{\lnot\RG}$ ($\vec{\eta}\equiv\vec{\varepsilon}$).  So, the energy functional (\ref{rgci:varrg:functional}) becomes a function of a single parameter 
\begin{equation}\label{rgci:varrg:energyfunction}
E[g]=\left.E[\vec{\eta},\vec{x}(\vec{\eta},g)]\right|_{\vec{\eta}\equiv\vec{\varepsilon}},
\end{equation}
and the variational procedure reduces to finding that particular interaction strength $g_0$ that minimizes the energy function
\begin{equation}\label{rgci:varrg:optimalg}
g_0=\arg\min_g E[g].
\end{equation}
The benefits of this major simplification are (a) the elimination of any ambiguity in the single-particle parameters $\vec{\eta}$ in the RG model, (b) the reduction in computational cost from a gradient descent method to a single-parameter line search, and (c) the possibility of a quick visual assessment of the optimal solution.  These advantages come at the price of a smaller variational space, and a corresponding reduction in correlation energy recovered in the optimized state.  However, it can be anticipated that the loss in correlation energy will be quickly recovered in the subsequent CI step in the RGCI approach.

\begin{figure}[!htb]
	\includegraphics{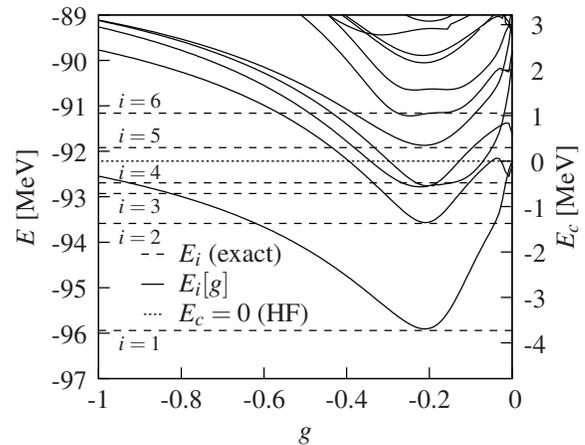}
	\caption{Full lines represent the energy function profiles $E_i[g]$ (\ref{rgci:varrg:functional}) for different eigenstates $i$ of the corresponding RG Hamiltonian (\ref{rg:hamiltonian}).  Exact eigenstate energies $E_i$(exact) of the effective Hamiltonian (\ref{rgci:hamiltonian}) are given in dashed lines ($i=1\dots6$).  The energy scale shows total (left axis) and correlation energies (right axis).   The HF reference energy for the correlation energies is shown as a dotted line.  The calculations are performed for $N=8$ pairs (${}^{116}$Sn).}\label{figure:rgci:varrg:116sn}
\end{figure}

Results of the variational calculation for $N=8$ pairs (${}^{116}$Sn) are presented in Figure \ref{figure:rgci:varrg:116sn}.
The exact ground-state energy for this isotope with the $G$-matrix Hamiltonian (\ref{rgci:hamiltonian}) is $E=-95.942$MeV, corresponding to a correlation energy of $E_c=-3.728$MeV.  The correlation energy is defined as the ground-state energy, corrected by the Hartree-Fock energy $E_{\textrm{HF}}$, obtained by filling the $N$ lowest single-particle energy levels up to the Fermi level.  In our case, the latter corresponds exactly to the energy function (\ref{rgci:varrg:energyfunction}) evaluated at $g=0$
\begin{equation}\label{rgci:varrg:correlationenergy}
E_c=E-E_{\textrm{HF}}=E-E[g=0].
\end{equation}
For easy comparison, both the total energy scale (left axis) as well as the correlation energy scale (\ref{rgci:varrg:correlationenergy}) (right axis) are present in the Figure.  The dashed lines are the exact reference energies for the first six eigenstates of the Hamiltonian (\ref{rgci:hamiltonian}) ($E_i$(exact), $i=1\dots 6$), and the  dotted line represents the HF energy, or zero correlation energy $E_c=0$ value.  The full lines are the values of the energy functional (\ref{rgci:varrg:energyfunction}) for different eigenstates of the corresponding RG Hamiltonian, as a function of $g$.  The lowest full curve in the Figure corresponds to the energy functional of the RG ground state, and gives the best approximation of the exact ground-state energy, as was expected.  The variationally obtained energy is reached at $g_0=-0.211$MeV, giving rise to $E[g_0]=-95.907$ MeV, which is equivalent to $99.07\%$ of the exact correlation energy.  A more practical measure to gauge the quality of a method is given by 1 minus this correlation-energy ratio, being
\begin{equation}\label{rgci:varrg:deltac}
\delta_c=1-\frac{E_c(\textrm{method})}{E_c(\textrm{exact})},
\end{equation}
which in the present example amounts to $0.93\%$.  This is a promising starting point for the RGCI method, certainly in light of the goals set in recently developed similar methods \cite{degroote:2016,ripoche:2017}, which have reported results around $1\%$ and lower.  

Apart from the RG ground-state energy curve, it is also interesting to investigate the performance of other RG eigenstates energy curves.  These are also included in Figure \ref{figure:rgci:varrg:116sn}.  It is clear that the low-lying excited RG energy curves $E_i[g]$ all approach an exact eigenstate energy $E_i$(exact) in the vicinity of the optimal $g_0=-0.211$MeV, pointing out that the integrable RG Hamiltonian (\ref{rg:hamiltonian}) with $\vec{\eta}=\vec{\varepsilon}$ and $g=g_0=-0.211$MeV is indeed a good approximation to the effective Hamiltonian (\ref{rgci:hamiltonian}).  It is worth pointing out that $g_0$ almost coincides with the average pairing interaction strength 
\begin{equation}\label{rgci:varrg:meanv}
\bar{V}=\frac{\sum_{ik}\Omega_iV_{ik}\Omega_k}{\sum_{ik}\Omega_i\Omega_k} =-0.212 \textrm{MeV},
\end{equation}
with the pairing interaction matrix elements $V_{ik}$ and degeneracies listed in Table \ref{table:interactionparameters}.  

\begin{figure}[!htb]
	\includegraphics{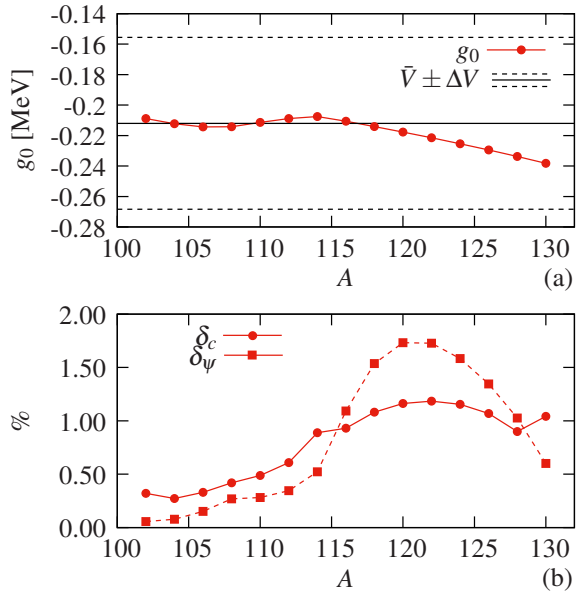}
	\caption{Upper panel (a) depicts values of $g_0$ (red dots) for all isotopes, compared to the mean interaction strength $\bar{V}=-0.212$MeV (full line).  The standard deviation ($\Delta V$) on the realistic interaction is represented as an error band (dotted lines).  Lower panel (b) gives quality measures $\delta_c$ (dots) (\ref{rgci:varrg:deltac}) and $\delta_\psi$ (squares) (\ref{rgci:varrg:deltapsi}) of the varRG method for different isotopes.}\label{figure:rgci:varrg:g0deltacpsi}
\end{figure}

Similar results are obtained for the other isotopes in the shell.  An overview of the optimal values $g_0$, and the corresponding missing correlations energies $\delta_c$ can be found in Figures \ref{figure:rgci:varrg:g0deltacpsi}(a) and \ref{figure:rgci:varrg:g0deltacpsi}(b) respectively, while numerical values are listed in Table \ref{table:rgci:overview}.
\begin{table}[!htb]
\begin{tabular}{lr|r|cr|cr|r}
${}^A$Sn & $N$ & $\dim\mathcal{H}$ & $g_0$ [MeV] & $\delta_c$ [\%] & $g_b$ [MeV]& $\delta_c$ [\%] & $b_b/b_0$\\
\hline
102 & 1 &   5 & -0.209 & 0.32 & -0.226 &  0.72 & 2.78\\
104 & 2 &  14 & -0.212 & 0.27 & -0.294 &  7.78 & 1.91\\
106 & 3 &  29 & -0.214 & 0.33 & -0.277 &  5.65 & 1.48\\
108 & 4 &  49 & -0.214 & 0.41 & -0.249 &  2.48 & 1.31\\
110 & 5 &  71 & -0.211 & 0.48 & -0.274 &  7.85 & 1.17\\
112 & 6 &  91 & -0.209 & 0.61 & -0.302 & 17.61 & 1.29\\
114 & 7 & 105 & -0.208 & 0.88 & -0.327 & 30.27 & 1.34\\
116 & 8 & 110 & -0.211 & 0.93 & -0.317 & 17.17 & 1.35\\
118 & 9 & 105 & -0.214 & 1.08 & -0.306 &  9.92 & 1.40\\
120 & 10 & 91 & -0.218 & 1.16 & -0.299 &  6.22 & 1.19\\
122 & 11 & 71 & -0.222 & 1.18 & -0.248 &  1.73 & 1.14\\
124 & 12 & 49 & -0.225 & 1.11 & -0.264 &  1.94 & 1.28\\
126 & 13 & 29 & -0.230 & 1.06 & -0.265 &  1.57 & 1.17\\
128 & 14 & 14 & -0.234 & 0.90 & -0.374 &  4.29 & 1.47\\
130 & 15 &  5 & -0.238 & 1.04 & -0.249 &  1.07 & 1.03\\
\hline
\end{tabular}
\caption{Variationally optimal values $g_0$ (4th column) and the corresponding missing correlation energy error $\delta_c$ (5th column) for all isotopes in the $A=100-132$ shell of Sn.  The values $g_b$ (6th column) denote the RG basis for which convergence is fastest, its corresponding $\delta_c$ at the varRG level (7th column), and the ratio of the $b$ fitting parameter with respect to the variational optimum (8th column).}\label{table:rgci:overview}
\end{table}
Besides the difference in correlation energy $\delta_c$ (\ref{rgci:varrg:deltac}), the difference in overlap of the (normalized) Bethe Ansatz wavefunction at the optimal interaction strength $g_0$ with the exact ground state 
\begin{equation}\label{rgci:varrg:deltapsi}
\delta_\psi=1-\left|\frac{\langle\vec{\varepsilon},\vec{x}(\vec{\varepsilon},g_0)|\psi_{\textrm{exact}}\rangle}{\langle\vec{\varepsilon},\vec{x}(\vec{\varepsilon},g_0)|\vec{\varepsilon},\vec{x}(\vec{\varepsilon},g_0)\rangle}\right|^2,
\end{equation}
can also be calculated, and is depicted in Figure \ref{figure:rgci:varrg:g0deltacpsi}(b).  Both measures display a similar global behaviour, pointing out that the correlation errors are around the 1\% level for all isotopes under investigation.  However, as there is a clear correlation between $\delta_c$ and $\delta_\psi$ (see, e.g., the Appendix in \cite{gunst:2017}), we will only use the former as a quality measure in the present paper.  

The robustness of the varRG method is further illustrated in Figure \ref{figure:rgci:varrg:g0deltacpsi}(a) by comparing the optimal values $g_0$ with the mean value $\bar{V}$ of the interaction (\ref{rgci:varrg:meanv}).  In order to appreciate the small variance of $g_0$ with respect to $\bar{V}$, the standard deviation $\Delta V$ on the realistic interaction (in Table \ref{table:interactionparameters}) is also given.

\subsection{Richardson-Gaudin Configuration Interaction} 
In the next step, the variationally obtained RG state is employed as a starting point for constructing a Hilbert space $\mathcal{H}$ that is adapted to the non-integrable Hamiltonian $H_{\lnot\textrm{RG}}$ of interest (\ref{rgci:hamiltonian}).  The optimized interaction strength $g_0$ \emph{defines} an integrable RG Hamiltonian (\ref{rg:hamiltonian}), and therefore provides a complete Hilbert space of (on-shell) RG states (\ref{rg:state}), in which the matrix representation of (\ref{rgci:hamiltonian}) can be constructed.  So, the idea is to build a hierarchy of on-shell basis states and diagonalize the non-integrable Hamiltonian in an increasingly large basis until convergence is reached.   This is the key idea behind the Richardson-Gaudin Configuration Interaction (RGCI) method.  In essence, this step is equivalent to the Truncated Space Approximation (TSA) \cite{yurov:1990,yurov:1991,james:2017}, with the main difference that the basis has been pre-optimized in the present paper.

Several criteria to construct this hierarchy can be envisioned.  A natural choice is to start from the optimized RG state, and include excited states according to the energy expectation value $E[\vec{\eta},\vec{x}^{(i)}]$ (\ref{rgci:varrg:functional}) in the $i$th on-shell state $|\vec{\eta},\vec{x}^{(i)}\rangle$.  Note that the hierarchy label $(i)$ has been appended to the rapidities because each different on-shell state $|\vec{\eta},\vec{x}^{(i)}\rangle$ is uniquely characterized by a different solution $\vec{x}^{(i)}$ of the RG equations (\ref{rg:equations}).   This choice is a straightforward generalization of the common practice in conventional CI methods starting from a non-correlated Fock space.  However, in contrast to conventional Fock space CI, there are no readily available estimates of $\langle H_{\lnot\textrm{RG}}\rangle$, other than calculating the expectation value explicitly.  Following this logic, one would have to calculate the expectation value of all possible states in $\mathcal{H}$ to find the appropriate ranking of excited states.  This is not desirable, so we opt for a different criterion.  We fix the ordering of the on-shell basis states by means of the eigenstate energy spectrum of the RG Hamiltonian, and diagonalize the Hamiltonian $H_{\lnot\textrm{RG}}$ in an increasing active Hilbert space $\mathcal{H}_i$ ($i=1\dots \dim{\mathcal{H}}$) of on-shell states until convergence or the complete basis limit ($\mathcal{H}_{\dim\mathcal{H}}\equiv\mathcal{H}$) is reached.  

The steps in the RGCI procedure are then as follows
\begin{enumerate}
\item Choose a set of single-particle energies $\vec{\eta}$ and an interaction strength $g$ for the RG Hamiltonian $H_{\textrm{RG}}$ (\ref{rg:hamiltonian}).  In the present paper, we stick to $\vec{\eta}=\vec{\varepsilon}$ for simplicity, and take the variationally optimized $g_0$ (\ref{rgci:varrg:optimalg}).  To appreciate the performance of the variationally obtained basis, we also present results for other values of $g\neq g_0$.  
\item Construct the lowest-energy on-shell eigenstate $|\vec{\eta},\vec{x}^{(1)}\rangle$ of $H_{\textrm{RG}}$ (\ref{rg:hamiltonian}), and assign this state as the first state in the active Hilbert space $\mathcal{H}_i$ (at this point, we have $i=1$).
\item Evaluate the expectation energy $E[\vec{\eta},\vec{x}^{(1)}]$ (\ref{rgci:varrg:functional}).
\item Add one unit to $i$.  Construct the next excited-energy on-shell state $|\vec{\eta},\vec{x}^{(i)}\rangle$ of $H_{\textrm{RG}}$, and add this state to the active Hilbert space $\mathcal{H}_{i-1}$.  
\item Diagonalize the Hamiltonian $H_{\lnot\textrm{RG}}$ in the new active Hilbert space $\mathcal{H}_i$, and extract the ground state and ground-state energy
\item Reiterate steps 4.\ to 6.\ until convergence in the ground-state energy is reached.  In the present paper, we proceed until the full Hilbert space is exhausted to investigate the convergence. 
\end{enumerate}

\begin{figure}[!htb]
	\includegraphics{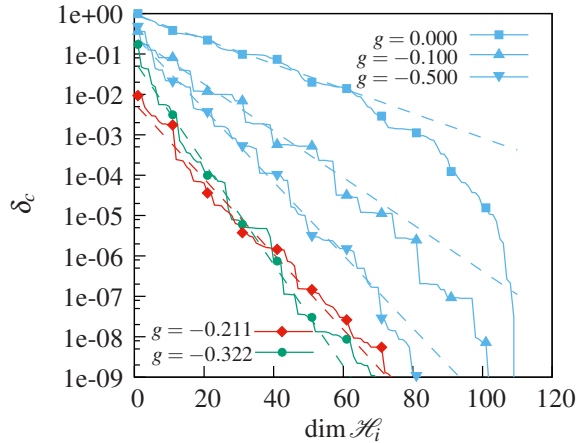}
	\caption{Convergence rate of missing correlation error $\delta_c$ for the RGCI method described for Hamiltonian with $N=8$ (${}^{116}$Sn).  Convergence rates for different integrable bases are denoted by the corresponding value of $g$.  Units of $g$ are given in MeV, and $\delta_c$ is dimensionless.}\label{figure:rgci:rgci:116sn}
\end{figure}

In Figure \ref{figure:rgci:rgci:116sn}, the convergence in the missing correlation energy error $\delta_c$ (\ref{rgci:varrg:deltac}) is presented for $N=8$ pairs (${}^{116}$Sn) for different values of $g$ (including $g_0$ in (red) diamonds).  
Obviously, $\delta_c$ is an adequate measure for the validation of the procedure, but only make sense when the exact ground state is known.  However, one can easily envision other suitable convergence measures in practical situations.

From Figure \ref{figure:rgci:rgci:116sn}, the following observations can be made.  
\begin{itemize}
\item Because of the variational principle, the error $\delta_c$ is monotically decreasing with increasing size of the active Hilbert space $\mathcal{H}_i$, and vanishes by definition as soon as the complete basis set limit is reached, regardless of the value of $g$.  For $N=8$ (${}^{116}$Sn), the complete basis limit is reached for $\dim\mathcal{H}=110$.  
\item Different values of $g$ give rise to different convergence rates.  The $g=0$ curve (blue squares) corresponds to the traditional approach in which the Hamiltonian $H_{\lnot\textrm{RG}}$ is diagonalised in an uncorrelated Fock space with increasing dimension.  As can be expected, the convergence rate of $\delta_c$ is steady but slow.  From Figure \ref{figure:rgci:rgci:116sn}, it can be seen that approximately half of the Hilbert space is required to build up the necessary degree of collectivity to reach the desired $\delta_c\le 1\%$ accuracy.  
\item For non-zero values of $g$, the convergence is considerably improved (note the log scale in the Figure).  This is visible in both the intercept and the slope of the $g\neq0$ curves.  Note that the values of the intercept correspond to the ground-state energy expectation value $E[g]$ (\ref{rgci:varrg:energyfunction}), so the more $g$ approaches the variational minimum $g_0$, the lower the value of the intercept.  The (red) curve with diamonds depicts exactly the RG basis constructed with the variationally optimized $g_0=-0.211$MeV.  Not only is the intercept lowest of all possible $g$ values by definition, the slope of convergence is also among the steepest, pointing out again that this is a very suitable basis.
\item It is palpable from the approximate linear behaviour of the curves in the log plot, that the convergence scales exponentially in the optimal cases.  To quantify this observation, an exponential fit of the form
\begin{equation}\label{rgci:rgci:fitexponential}
f(x) =\exp(a+bx),
\end{equation}
with $x=\dim\mathcal{H}_i$ is performed for each of the curves.  The parameters $a$ and $b$ account for the intercept and slope respectively of the curves in the log plot of Figure \ref{figure:rgci:rgci:116sn}.  The error loss function $\chi^2$ can be tailored such that it highlights the relevant features of the method, \emph{i.e.}\ the intercept and global convergence rate for low-dimensional active Hilbert spaces ($\dim\mathcal{H}_i\ll\dim\mathcal{H}$).  So, the used loss function is
\begin{equation}\label{rgci:rgci:chi2}
\chi^2=\sum_{x=1}^{\dim\mathcal{H}/2}|\ln \delta_c(x)-(a+bx)|^2,
\end{equation}
which manages to focus on the global convergence rate for small active Hilbert spaces ($\dim\mathcal{H}_i\le\dim\mathcal{H}/2$).  Note that in this case, the fitting procedure becomes a standard linear fitting problem.   A plot of the intercept $a$ and slope $b$ parameters for a range of $g$ values for $N=8$ (${}^{116}$Sn) is given in Figure \ref{figure:rgci:rgci:fit}.
\begin{figure}[!htb]
	\includegraphics{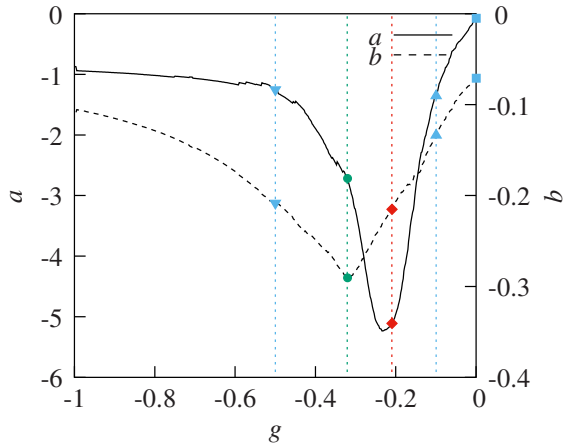}
	\caption{Fitting parameters $a$ (solid line) and $b$ (dashed line) of fitting function $f(x)$ (\ref{rgci:rgci:fitexponential}) for a range of basis sets, corresponding to different values of $g$.  The highlighted values of $g$ correspond to the presented curves in Figure \ref{figure:rgci:rgci:116sn}.}\label{figure:rgci:rgci:fit}
\end{figure}
The qualitative behaviour of both parameters confirms the results from the calculations.  First, the intercept $a$ is indeed minimal around the variationally obtained value $g_0$ (\ref{rgci:varrg:optimalg}).  Note that the lowest value $a$ does not occur exactly at $g_0=-0.211$MeV.  However, this is due to details in the definition of the loss function (\ref{rgci:rgci:chi2}).  Second, the slope parameter $b$ follows a similar behaviour as the intercept $a$, pointing out that the convergence rate is indeed quicker around the optimal value $g_0$.  Interestingly, the fastest convergence is not reached at $g_0=-0.211$MeV, but a little bit further ($g=-0.322$MeV).  This can be verified in Figure \ref{figure:rgci:rgci:116sn}, where the (green) $g=-0.322$MeV tumbles below the (red) $g_0=-0.211$MeV curve at around $\dim\mathcal{H}_i=40$.  However, it should be kept in mind that the $g=-0.322$MeV starts from a suboptimal $\delta_c$ at $i=1$, and only becomes significantly better at larger dimensions of the active Hilbert space.  Nevertheless, this observation may point out that a different hierarchy of basis states may lead to a further optimization of the missing correlation error.
\end{itemize}

\begin{figure*}[!htb]
\includegraphics{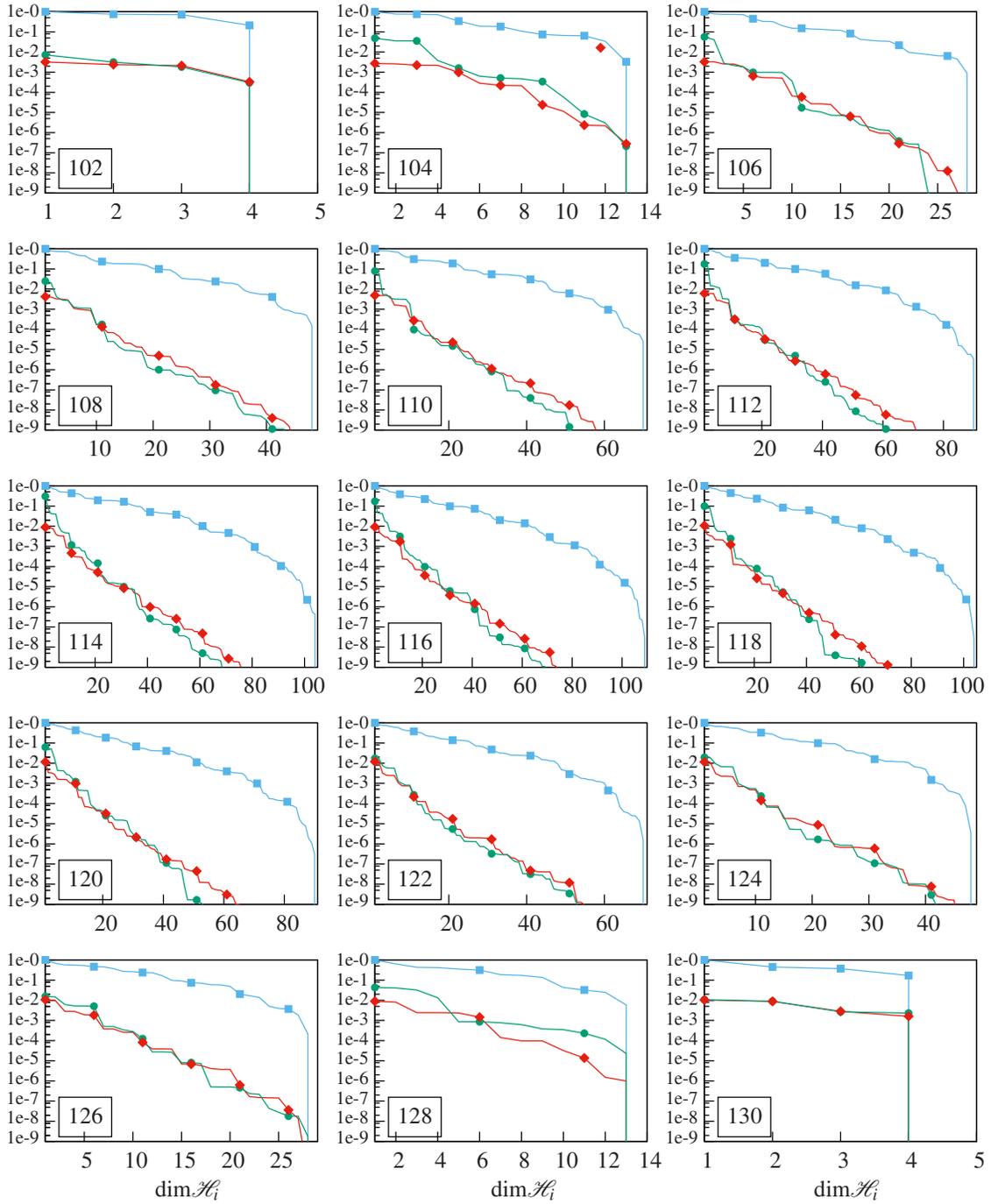}
\caption{Convergence rate of missing correlation error $\delta_c$ for the RGCI method for different istopes of ${^{A}}$Sn (with $A$ in lower left corner of each panel).  Convergence rates for different integrable bases are denoted by the corresponding value of $g$.  (Blue) squares correspond to $g=0.000$MeV, (red) diamonds to the variationally optimized $g_0$, and (green) dots to the fastest converging basis.  Units of $g$ are given in MeV, and $\delta_c$ is dimensionless.}\label{fig:rgci:rgci:isotopes}
\end{figure*}

The results presented for $N=8$ are generic for all isotopes, as shown in Figure \ref{fig:rgci:rgci:isotopes}.  
Each panel in this figure contains the curves for the variationally optimized $g_0$ basis (red diamonds), the fastest converging basis (green dots), and uncorrelated basis $g=0.000$MeV (blue squares) for the denoted isotope (lower left corner of each panel).  The explicit numerical values of $g_b$ are listed in Table \ref{table:rgci:overview}.  Note that panel 116 in Figure \ref{fig:rgci:rgci:isotopes} shows selected results from Figure \ref{figure:rgci:rgci:116sn}, for comparison.  The convergence rate of $\delta_c$ is consistently faster for all isotopes when the variationally optimized $g_0$ is chosen for the on-shell basis compared to the conventional Fock space $g=0$.  For each isotope, there exists a basis where the convergence is quicker (green dots), however this happens in most cases at higher dimensions of the active Hilbert space (see Table \ref{table:rgci:overview}).  The exceptions to the general observation are for the isotopes ${}^{102}$Sn and  ${}^{130}$Sn, corresponding to respectively one particle pair ($N=1)$ and one hole pair ($N=15$), where the improvements from the extra CI step of the RGCI is negligible with respect to the variational optimization.  The naive understanding of this result is that all collective features of the $N=1$ (Cooper) pair state have been captured by the integrable model, and that all possible corrections necessarily come from non-collective excited states.  Opening up the single-particle channels $\vec{\eta}$ as variational parameters in the functional (\ref{rgci:varrg:functional}), as opposed to fixing it as $\vec{\eta}=\vec{\varepsilon}$, can incorporate these corrections exactly for $N=1$ by construction.  This is because the number of variational parameters then matches the size of the Hilbert space.

\subsection{Pre-diagonalization and Similarity Renormalization Group}
\begin{figure}[!htb]
	\includegraphics{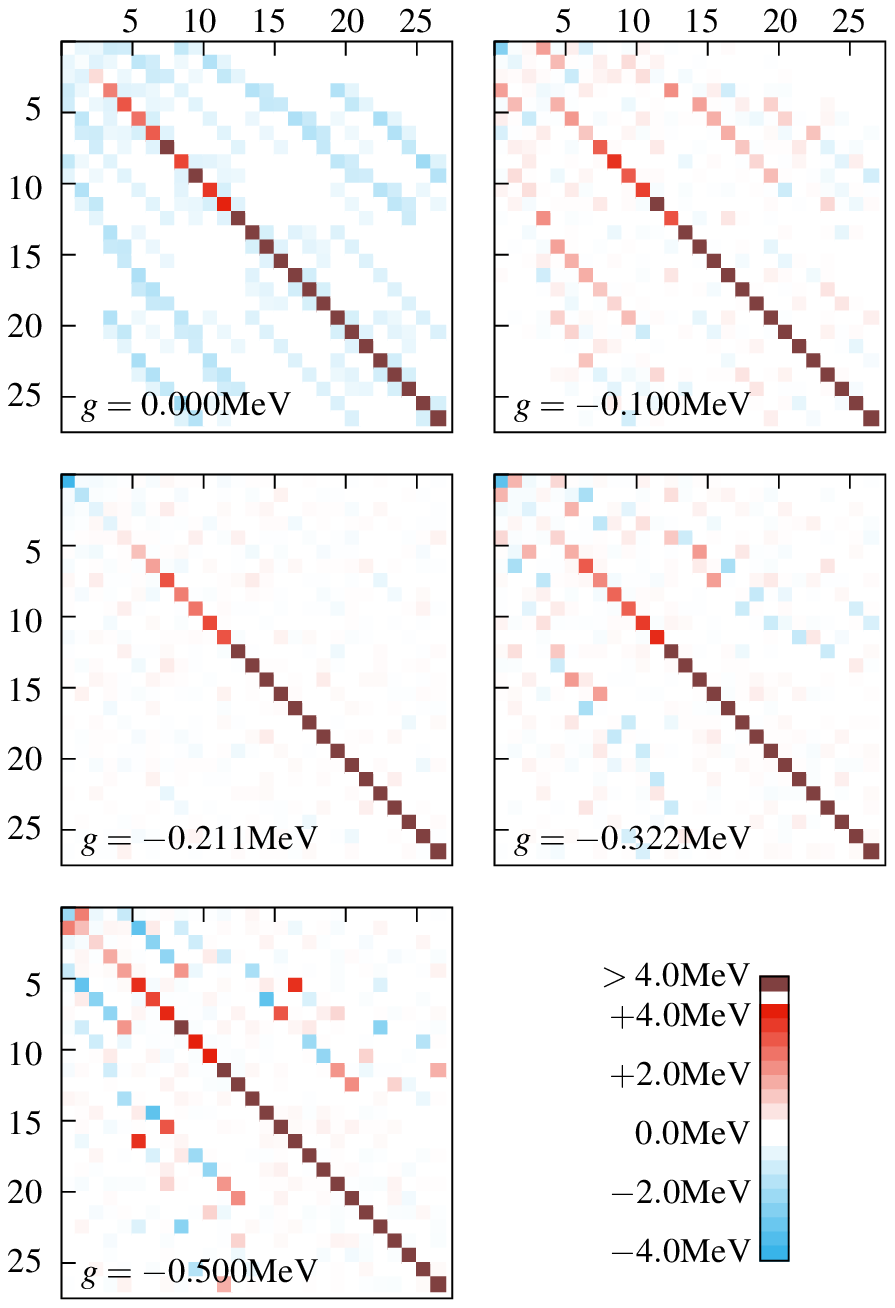}
	\caption{Visual representation the exact Hamiltonian $H_{\lnot\textrm{RG}}$ (\ref{rgci:hamiltonian}) matrix in different normalized Richardson-Gaudin bases, labeled by $g$ (in MeV).  Each square represents a matrix element (\ref{rgci:srg:matrixelement}), with the value of the shading denoting the magnitude of the matrix element.  The sign of the matrix element (blue/red color) is irrelevant for the discussion.  Only the lowest (quarter) part of the Hamiltonian matrix is shown. }\label{figure:rgci:srg:matrixelements}
\end{figure}

Although intuitive, the good convergence rate of the RGCI method at the variational minimum is by no means guaranteed from the variational principle.  For a better understanding of the convergence performance of RGCI, it is instructive to investigate the matrix elements of the non-integrable Hamiltonian (\ref{rgci:hamiltonian}) in the basis of on-shell RG states (\ref{rg:state}) 
\begin{equation}\label{rgci:srg:matrixelement}
\frac{\langle\vec{\varepsilon},\vec{x}(g)|H_{\lnot\textrm{RG}}|\vec{\varepsilon},\vec{y}(g)\rangle}{\sqrt{\langle\vec{\varepsilon},\vec{x}(g)|\vec{\varepsilon},\vec{x}(g)\rangle\langle\vec{\varepsilon},\vec{y}(g)|\vec{\varepsilon},\vec{y}(g)\rangle }},
\end{equation}
as a function of $g$.  These matrix elements are visualized in Figure (\ref{figure:rgci:srg:matrixelements}) 
for $N=8$ (${}^{116}$Sn) with the same selected values of $g$ as in Figures \ref{figure:rgci:rgci:116sn} and \ref{figure:rgci:rgci:fit}.  For visual purposes, only the lowest part of the total matrix is given.  Each matrix element is represented by a colored dot, with the  color saturation proportional to the magnitude of the matrix elements (\ref{rgci:srg:matrixelement}).  The Figure distinguishes between positive and negative matrix elements, however this distinction is irrelevant as particular matrix elements can be sign flipped by an appropriate phase similarity transformation.  More important, zero-valued matrix elements are represented by white dots.  The diagonal matrix elements are shifted such that the first matrix element (upper left dot in each panel) represents the correlation energy $E_c$ (see eq.\ (\ref{rgci:varrg:correlationenergy}) and right axis of Figure \ref{figure:rgci:varrg:116sn}).  The $g=0.000$MeV panel corresponds to the traditional Hamiltonian matrix in Fock space.  Accordingly, the upper-left matrix element is zero (white), by definition.  Moving away from $g=0.000$MeV, the off-diagonal elements of the matrix become suppressed, with the Hamiltonian matrix (\ref{rgci:srg:matrixelement}) approaching diagonality around the variationally optimal value $g_0=-0.211$MeV.  It is worth noting that the diagonality is again lost when further increasing $g$, even at the fastest converging point $g_b=-0.322$MeV.  From this, it is easy to understand the fast convergence of the RGCI method at the optimal varRG state, as the Hamiltonian matrix was already very close to diagonal from the start.

This observation appears to be in line with ideas from Similarity Renormalization Group (SRG) methods \cite{wegner:1994,bogner:2010,hergert:2017}.  The SRG describes an isospectral flow of a Hamiltonian in such a way that it finds a representation (basis) in which part of the Hamiltonian matrix is suppressed.  The varRG method shares the characteristics of an isospectral flow because the Hamiltonian matrix (\ref{rgci:srg:matrixelement}) can be recast as a unitary similarity transformation
\begin{align}
&\frac{\langle\vec{\varepsilon},\vec{x}(g)|H_{\lnot\textrm{RG}}|\vec{\varepsilon},\vec{y}(g)\rangle}{\sqrt{\langle\vec{\varepsilon},\vec{x}(g)|\vec{\varepsilon},\vec{x}(g)\rangle\langle\vec{\varepsilon},\vec{y}(g)|\vec{\varepsilon},\vec{y}(g)\rangle }}\notag\\
&=\sum_{\vec{n},\vec{n}^\prime}\frac{\langle\vec{\varepsilon},\vec{x}(g)|\vec{n}\rangle\langle\vec{n}|H_{\lnot\textrm{RG}}|\vec{n}^\prime\rangle\langle\vec{n}^\prime|\vec{\varepsilon},\vec{y}(g)\rangle}{\sqrt{\langle\vec{\varepsilon},\vec{x}(g)|\vec{\varepsilon},\vec{x}(g)\rangle\langle\vec{\varepsilon},\vec{y}(g)|\vec{\varepsilon},\vec{y}(g)\rangle }},
\end{align}
with $\{|\vec{n}\rangle\}$ and $\{|\vec{n}^\prime\rangle\}$ both a complete set of (normalized) basis states in Fock space.  In operator form, this can be clarified as
\begin{equation}
H_{\lnot\textrm{RG}}(g)=U(g)H_{\lnot\textrm{RG}}(0)U(g)^\dag,
\end{equation}
with $H_{\lnot\textrm{RG}}(0)$ the matrix representation in Fock space, and $U(g)$ the unitary matrix with matrix elements
\begin{equation}\label{rgci:srg:unitary}
U(g)_{\vec{x},\vec{n}}=\frac{\langle\vec{\varepsilon},\vec{x}(g)|\vec{n}\rangle}{\sqrt{\langle\vec{\varepsilon},\vec{x}(g)|\vec{\varepsilon},\vec{x}(g)\rangle}}.
\end{equation}
The varRG method then shares the properties of isospectral flow with SRG because each value of $g$ not only characterizes a (variational) trial state, but also a complete basis of on-shell Bethe Ansatz states, leading to a full-rank unitary matrix (\ref{rgci:srg:unitary}).  This is in contrast with other variational approaches, where typically only the trial state is properly defined.  Nevertheless, the main difference with SRG is that SRG generates a dynamical flow from local updates driven towards a suppression of unwanted off-diagonal matrix elements.  In varRG, the suppression of the off-diagonal part of the Hamiltonian matrix appears to be a convenient byproduct of the variational approach leading to optimal convergence properties in the RGCI step. 
\subsection{Correlation Coefficients}
Closely related to the missing overlap $\delta_\psi$ (\ref{rgci:varrg:deltapsi}), the deviations from the exact correlation coefficients
\begin{align}
\Pi_{ik}=\langle\psi_\textrm{exact}|\hat{S}_i^\dag \hat{S}_k|\psi_\textrm{exact}\rangle,\label{rgci:rgci:corcoef:pi}\\
D_{ik}=\langle\psi_\textrm{exact}|\hat{n}_i\hat{n}_k|\psi_\textrm{exact}\rangle,
\end{align}
provide a detailed measure to gauge the performance of an approximation method because they are more sensitive to details in the structure than simple energy measures.  So, for completenes, the values for $\Pi_{ik}=\langle\psi|S_i^\dag S_k|\psi\rangle$ (\ref{rgci:rgci:corcoef:pi}) are given in Table \ref{table:rgci:corcoef} for $g=0.000$MeV, and $g_0=-0.211$MeV at the varRG level, and the exact values for $N=8$ (${}^{116}$Sn).  
\begin{table}[!htb]
\begin{tabular}{r|rrrrr}
\hline
exact & $g_{\frac{7}{2}}$ & $d_{\frac{5}{2}}$ & $s_{\frac{1}{2}}$ & $h_{\frac{11}{2}}$ & $d_{\frac{3}{2}}$ \\
\hline
            $g_{\frac{7}{2}}$  &   4.737 &  1.523 &  0.644 &  3.630 &  1.193\\
            $d_{\frac{5}{2}}$  &         &  3.393 &  0.637 &  3.427 &  1.231\\
            $s_{\frac{1}{2}}$  &         &        &  0.323 &  1.243 &  0.431\\
            $h_{\frac{11}{2}}$ &         &        &        &  6.502 &  2.169\\
            $d_{\frac{3}{2}}$  &         &        &        &        &  0.840\\
\hline\hline
$g=-0.211$ & $g_{\frac{7}{2}}$ & $d_{\frac{5}{2}}$ & $s_{\frac{1}{2}}$ & $h_{\frac{11}{2}}$ & $d_{\frac{3}{2}}$ \\
\hline
            $g_{\frac{7}{2}}$  &   4.773 &  1.543 &  0.642 &  3.734 &  1.234\\
            $d_{\frac{5}{2}}$  &         &  3.389 &  0.597 &  3.459 &  1.142\\
            $s_{\frac{1}{2}}$  &         &        &  0.252 &  1.166 &  0.383\\
            $h_{\frac{11}{2}}$ &         &        &        &  6.910 &  2.198\\
            $d_{\frac{3}{2}}$  &         &        &        &        &  0.798\\
\hline\hline
$g=0.000$ & $g_{\frac{7}{2}}$ & $d_{\frac{5}{2}}$ & $s_{\frac{1}{2}}$ & $h_{\frac{11}{2}}$ & $d_{\frac{3}{2}}$ \\
\hline
            $g_{\frac{7}{2}}$  &   4.000 &  0.000 &  0.000 &  0.000 &  0.000\\
            $d_{\frac{5}{2}}$  &         &  3.000 &  0.000 &  0.000 &  0.000\\
            $s_{\frac{1}{2}}$  &         &        &  1.000 &  0.000 &  0.000\\
            $h_{\frac{11}{2}}$ &         &        &        &  0.000 &  0.000\\
            $d_{\frac{3}{2}}$  &         &        &        &        &  0.000\\
\hline\hline
\end{tabular}
\caption{Correlation coefficients $\Pi_{ik}=\langle\psi|\hat{S}_i^\dag \hat{S}_k|\psi\rangle$ (\ref{rgci:rgci:corcoef:pi}) for the exact wavefunction (upper table), the uncorrelated $g=0.000$ Fock basis (lower table) and the variationally optimized $g=-0.211$ (middle table) at the varRG level.  Energies are given in MeV, and correlation coefficients are dimensionless.}\label{table:rgci:corcoef}
\end{table}
The values in the table are consistent with the other results throughout the paper.  The deviations of the correlation coefficients at the variationally optimized $g_0$ are typically within the 1 to 10\% range, as opposed to the conventional Fock space basis, where the matrix elements are even qualitatively wrong.  Moving into the RGCI step again induces an exponentially fast convergence to the exact values around $g_0$ (not shown).  
\section{Conclusions and Outlook}
We have presented a new method for the treatment of pairing correlations.  The method consists of two consecutive steps.  The first step is a variational optimization of an on-shell Richardson-Gaudin state, as pioneered recently for quantum chemistry \cite{tecmer:2014,johnson:2015} and quantum dots \cite{claeys:2017b}.  The benefits of using a Richardson-Gaudin ground state for nuclear structure physics is that this wave function is already qualitatively correct for the description of collective Cooper pair condensation.  This eliminates the need for a sophisticated selection scheme to identify the correct manifold of on-shell states upon which to vary.  The second step is to use the resulting set of excited states on top of the variationally optimized Richardson-Gaudin state as a basis in which to perform a Configuration Interaction calculation in an increasingly large active Hilbert space until convergence.  This is possible at computationally soft (polynomial) scaling, by virtue of the Slavnov theorem of integrability \cite{zhou:2002,faribault:2008,claeys:2015a,claeys:2017a}.  Again, the integrability of the Richardson-Gaudin model is key for the feasibility of this step.  Interestingly, the convergence to the exact values is exponential, mainly due to a strong suppression of the off-diagonal matrix elements in the Hamiltonian when expressed in this optimized basis.

In the present paper, the method has been confronted with a realistic pairing interaction, obtained from a $G$-matrix formalism for the Sn isotopes \cite{holt:1998,zelevinsky:2003}.  This interaction has been constructed for pure pairing correlations only.  It is well-known that nuclear structure physics consists of a competition between pairing and quadrupole correlations.  The future challenge will be to include quadrupole correlations in the present scheme.  There are a few tentative solutions for this.  One solution is to work in a deformed Nilsson basis instead of the spherical basis used in this paper \cite{nilsson:1995}, and project on good angular momentum states after the varRG and/or RGCI step.  Another approach would be to enlarge the RG basis set to include non-zero seniority states.  In the theory of Richardson-Gaudin integrability, this corresponds to the simple blocking of a given orbital, so all useful features of integrability for the varRG/RGCI method are kept.  A final approach would be to generalize the Slavnov-like theorems of integrability to higher-order algebras, like the isovector/scalar proton-neutron pairing algebras $so(5)$ and $so(8)$ \cite{dukelsky:2006,lerma:2007}, or the symplectic $sp(3,\mathbb{R})$ \cite{rosensteel:1977}.  However, much more mathematical results are needed for the efficient calculation of off-diagonal matrix elements \cite{beliard:2013,johnson:2013,johnson:2017}, so the first two suggestions seem much more straightforward in the short run.  From a physical point of view, it would be interesting to further investigate the connection between the varRG method and Similarity Renormalization Group \cite{hergert:2017} ideas.  Also, the applicability of the RGCI in other domains of physics is worth exploring.  For instance, it would be interesting to explore variationally optimized basis sets in the TSA.  Another domain of applicability is quantum chemistry, where the RGCI can provide a natural framework to extend variational geminal theory \cite{tecmer:2014,johnson:2015}.    
\begin{acknowledgements}
The authors acknowledge illuminating discussions with and support from Patrick Bultinck, Kris Heyde, Frank Verstraete (Universiteit Gent), Peter Limacher (Karlsruher Institut f\"ur Technologie), Paul Johnson (Universit\'e Laval), and  Veerle Hellemans (Vrije Universiteit Brussel).   We thank Neil J.\ Robinson (Universiteit van Amsterdam) for pointing out the connection between the Truncated Space Approximation and the Configuration Interaction part of the RGCI.  SDB acknowledges Thomas Duguet for his hospitality during an intensive lecture week at CEA Saclay.  PWC acknowledges support from a PhD fellowship and a travel grant for a long stay abroad at the University of Amsterdam from the Research Foundation Flanders (FWO-Vlaanderen). SDB and DVN acknowledge financial support from FWO Vlaanderen. J-SC acknowledges support from the Netherlands Organization for Scientific Research (NWO), and from the European Research Council under ERC Advanced grant 743032 DYNAMINT. This work is part of the Delta ITP consortium, a program of the Netherlands Organisation for Scientific Research (NWO) that is funded by the Dutch Ministry of Education, Culture and Science (OCW).
\end{acknowledgements}

\end{document}